\documentstyle[12pt]{article}

\global\arraycolsep=1pt
\newcommand{\resection}[1]{\setcounter{equation}{0}\section{#1}}

\newcommand{\bel}[1]{\begin{equation}\label{#1}}                     
\newcommand{\bal}[1]{\begin{eqnarray}\label{#1}}                     
\newcommand{\Eq}[1]{(\ref{#1})}                                    
\newcommand{\be}{\begin{equation}}
\newcommand{\ee}{\end{equation}}
\newcommand{\ba}{\begin{eqnarray}}
\newcommand{\ea}{\end{eqnarray}}
\newcommand{\nn}{\nonumber \\}

\newcommand{\vac}{{\rm vac}}

\newcommand{\co}{{\cal O}}
\newcommand{\dis}{\displaystyle}
\newcommand{\scr}{\scriptstyle}
\newcommand{\qq}{\qquad}

\begin{document}

\begin{titlepage}

\begin{flushleft}
{\baselineskip=14pt 
\rightline{
 \vbox{\hbox{YITP-98-22}
%       \vskip 2mm
       \hbox{April 1998}       }}}
\end{flushleft}

\vskip 2.0 cm

\begin{center}

%%%%%%%%%%%%%%%%%%%%%%%%%%%%%%%%%%%%%%%%%%%%%%%%%%%%%%%%%%%%%%%%%%%%%%
%%
%%   Title:
%%
{\large\bf Two-point correlation functions 
            in perturbed minimal models} \\
\vskip 3.0cm

%%%%%%%%%%%%%%%%%%%%%%%%%%%%%%%%%%%%%%%%%%%%%%%%%%%%%%%%%%%%%%%%%%%%%%%%
%%
%% Authors:
%%
{Takeshi Oota}
\footnote{e-mail: {\tt toota@yukawa.kyoto-u.ac.jp}}\\[2em]

%%%%%%%%%%%%%%%%%%%%%%%%%%%%%%%%%%%%%%%%%%%%%%%%%%%%%%%%%%%%%%%%%%%%%%%%
%%
%% Addresses:
%%
{\sl Yukawa Institute for Theoretical Physics, Kyoto
           University, Kyoto 606-01, Japan}\\

\end{center}

\vskip24pt

%%%%%%%%%%%%%%%%%%%%%%%%%%%%%%%%%%%%%%%%%%%%%%%%%%%%%%%%%%%%%%%%%%%%%%%
%%
%%   Abstract:
%%
\begin{abstract}
\noindent
Two point correlation functions of the off-critical primary fields
$\phi_{1, 1+s}$ are considered in the perturbed minimal models
$M_{2, 2N+3} + \phi_{1,3}$. They are given
as infinite series of form factor contributions. 
The form factors of $\phi_{1, 1+s}$ are conjectured from the known
results for those of $\phi_{1,2}$ and $\phi_{1,3}$. The conjectured 
form factors are rewritten in the form which
is convenient for summing up. The final expression of the two point
functions is written as a determinant of an integral operator.
\end{abstract}

\end{titlepage}

%%%%%%%%%%%%%%%%%%%%%%%%%%%%%%%%%%%%%%%%%%%%%%%%%%%%%%%%%%%%%%%%%%%%%%%%

\resection{Introduction}

Correlation functions are important tools to study quantum field
theories. In many two-dimensional models, 
it is known that the determinant representation is useful for
non-perturbative analysis of correlation functions 
\cite{BMW}--\cite{LLSS}.

In a class of $1+1$ dimensional, massive, integrable
models \cite{BKW}--\cite{L}, 
correlation functions of some operators can be
written as an infinite sum over intermediate states and are
analysed through the form factor bootstrap procedure
\cite{BKW,Sm}.

Recently, it is shown that determinant representation of
integral operators is useful to sum up the infinite series in the 
sinh-Gordon model \cite{KS} and in the scaling Lee-Yang model
\cite{KO}. In these models, an auxiliary Fock space and auxiliary Bose 
fields, which are called dual fields, are introduced. This approach
was developed in \cite{KBI,Kor,Sl}

The scaling Lee-Yang model \cite{F} can be identified with 
the $N=1$ case of the perturbed
minimal model $M_{2, 2N+3}+\phi_{1,3}$ \cite{CM}.
The purposes of this paper are to generalize the result of \cite{KO} to
arbitrary $N$ and to show that the determinant
representation is useful also in the perturbed minimal conformal field 
theories. 

The minimal model $M_{2, 2N+3}$ is non-unitary and contains $N+1$
scalar primary fields $\phi_{1, 1+s}=\phi_{1, 2N+2-s}$ 
($s=0, \ldots, 2N+1$) with scaling dimensions 
$(\Delta_{(1, 1+s)}, \Delta_{(1, 1+s)})$ \cite{BPZ}:
\be
\Delta_{(1, 1+s)} = - \frac{ s(2N+1-s) }{ 2(2N+3) }, \qq
s=0, \ldots, 2N+1.
\ee
The primary operator $\phi_{1,1}=\phi_{1,2N+2}$ is the identity
operator. 

The $\phi_{1,3}$-perturbation of $M_{2, 2N+3}$ is known to be
integrable and is described by the $A_{2N}^{(2)}$-type factorizable
scattering theory. The mass spectrum of $A_{2N}^{(2)}$ theory
consists of $N$ scalar particles
with mass
\be
m_a = 2m \sin( a\pi /h), \qq a=1, \ldots, N,
\ee
where $h=2N+1$ is the Coxeter number of the Lie algebra
$A_{2N}^{(2)}$. The two-body scattering amplitude is given by 
\cite{FKM}
\bel{Sab}
S_{ab}(\beta) = \prod_{ \scr x = |a-b|+1 \atop\scr {\rm step} 2}^{a+b-1}
\{ x \}_{(\beta)},
\ee
where
\be
\{ x \}_{(\beta)} = 
\frac{
\tanh\frac{1}{2}(\beta+(x-1)\pi i/h)
\tanh\frac{1}{2}(\beta+(x+1)\pi i/h)
}{
\tanh\frac{1}{2}(\beta-(x-1)\pi i/h)
\tanh\frac{1}{2}(\beta-(x+1)\pi i/h)
}.
\ee
It is conjectured that the conformal primary fields $\phi_{1, 1+s}$
become off-critical primary fields \cite{K}. 
We use same symbol $\phi_{1, 1+s}$
to denote the corresponding off-critical primary operators.

Form factors of a local operator $\co(x)$
are defined as the matrix elements between the vacuum state 
$\langle \vac |$ and $n$ particle states characterized by particle
species $a_i$ ($a_i\in \{ 1, \ldots, N\}$) and rapidities $\beta_i$
( $i= 1, \ldots, n$ ):
\be
F^{\co}_{a_1 \ldots a_n}
(\beta_1, \ldots, \beta_n) =
\langle \vac | \co(0) | \beta_1, \ldots, \beta_n 
\rangle_{a_1 \ldots a_n}.
\ee
The multi-particle form factors for $\phi_{1, 2}$ and $\phi_{1,3}$ 
were calculated in \cite{S}
\bel{phi12}
F_{a_1 \ldots a_n}^{\phi_{1,2}}(\beta_1, \ldots, \beta_n) =
f_{0; a_1 \ldots a_n}(\beta_1, \ldots, \beta_n)
\prod_{j=1}^n \nu_{a_j} \prod_{i<j}^n
\zeta_{a_i a_j}(\beta_i - \beta_j),
\ee
\bal{phi13}
& &F_{a_1 \ldots a_n}^{\phi_{1,3}}(\beta_1, \ldots, \beta_n) \nn
&=&
\frac{2\cos(\pi/h)}{m_1} \left(\sum_{j=1}^n m_{a_j} e^{\pm \beta_j} \right)
f_{\pm; a_1 \ldots a_n}(\beta_1, \ldots, \beta_n)
\prod_{j=1}^n \nu_{a_j} \prod_{i<j}^n
\zeta_{a_i a_j}(\beta_i - \beta_j).
\ea
The explicit forms of the constants $\nu_a$, the functions
$f_{\lambda; a_1 \ldots a_n}(\beta_1, \ldots, \beta_n)$ 
($\lambda = 0, \pm 1$) and $\zeta_{ab}(\beta)$ are given in section 2.1. 
Note that \Eq{phi13} gives two equivalent definitions of $\phi_{1, 3}$.

For other operators, the explicit form of multi-particle form factors
were determined only when $a_1 = \ldots = a_n = 1$ \cite{K}. 
The explicit form for the form factors containing the other particle
species had not been known.
These form factors used to be given indirectly by using the fusion
procedure. We will derive these in this paper.

After  Wick rotation to the Euclidean space, 
the two-point correlation function of the operator $\phi_{1, 1+s}$ and 
$\phi_{1, 1+s'}$ can be represented as an 
infinite series of form factors contributions
\bal{phiss}
& &\langle \phi_{1, 1+s}(x) \phi_{1, 1+s'}(0) \rangle \nn
&=&
\sum_{n=0}^{\infty} \sum_{a_i=1}^N \int \frac{d^n\beta}{n!(2\pi)^n}
\langle \vac| \phi_{1, 1+s}(x) |\beta_1, \ldots, \beta_n 
\rangle_{a_1 \ldots a_n} \ 
^{a_n \ldots a_1}\langle \beta_n, \ldots, \beta_1 | 
\phi_{1, 1+s'}(0) |\vac \rangle \nn
&=&
\sum_{n=0}^{\infty} \sum_{a_i} \int \frac{d^n\beta}{n!(2\pi)^n}
F_{a_1 \ldots a_n}^{\phi_{1, 1+s}}(\beta_1, \ldots, \beta_n)
F_{a_n \ldots a_1}^{\phi_{1, 1+s'}}(\beta_n + \pi i , \ldots, 
\beta_1 + \pi i) \nn
& & \qq \qq \qq \qq
\times \exp\left[-r \sum_{j=1}^n m_{a_j} \cosh \beta_j \right],
\ea
where $r = (x^{\mu} x_{\mu})^{1/2}$.
In the next section, we transform \Eq{phi12} and \Eq{phi13} to forms
which are convenient to sum up the series \Eq{phiss}.
From the final expression, we can guess the form of the form factors
for the other off-critical primary operators. 
We give the conjectured form factors for $\phi_{1, 1+s}$ and
demonstrate that they satisfy form factor bootstrap equations.
We discuss the relation between the conjectured form factors and 
their known forms with all $a_i=1$ given by Koubek \cite{K}. 

This paper is organized as follows. 
In the first part of section 2, a brief review of the form factor
bootstrap equations is given.
In section 2.1, the form factors
\Eq{phi12} and \Eq{phi13} are transformed to a form which is
convenient for summation.
In section 2.2, we give the form factors for 
other primary operators $\phi_{1, 1+s}$. In section 3, with the help
of dual fields which act on an auxiliary Fock space, we sum up the
infinite series \Eq{phiss} to a Fredholm determinant.
Section 4 is devoted to discussion.
In the appendix we give the evidences that the proposed form factors
of $\phi_{1, 1+s}$ satisfy the form factor bootstrap equations.

\resection{Form factor}

To fix a notation, we briefly summarize the form factor bootstrap 
equations \cite{BKW,Sm}.

The form factor bootstrap equations are axiomized in the
following way.

(i) Watson's equations:
\ba
& & F^{\co}_{a_1 \ldots a_i a_{i+1} \ldots a_n}(\beta_1, \ldots, \beta_i,
\beta_{i+1}, \ldots, \beta_n) \nn
&=& S_{a_i a_{i+1}}(\beta_i - \beta_{i+1})
F^{\co}_{a_1 \ldots a_{i+1} a_i \ldots a_n}
(\beta_1, \ldots, \beta_{i+1},
\beta_i, \ldots, \beta_n),
\ea
\be
F^{\co}_{a_1 a_2 \ldots a_n}
(\beta_1 + 2\pi i, \beta_2, \ldots, \beta_n) =
F^{\co}_{a_2 \ldots a_n a_1}(\beta_2, \ldots, \beta_n, \beta_1).
\ee
(ii) Lorentz covariance:
\be
F^{\co}_{a_1 \ldots a_n}(\beta_1 + \Lambda, \ldots, 
\beta_n + \Lambda ) = 
e^{s(\co)\Lambda}
F^{\co}_{a_1 \ldots a_n}(\beta_1, \ldots, \beta_n),
\ee
where $s(\co)$ is the Lorentz spin of the operator $\co$.
The off-critical primary fields are scalar operators : 
$s(\phi_{1, 1+s})=0$.

(iii) The kinematical residue equation:
\ba
& & -i \lim_{\epsilon \rightarrow 0} \epsilon
F^{\co}_{a a d_1 \ldots d_n}(\beta + \pi i + \epsilon, \beta,
\beta_1, \ldots, \beta_n) \nn
&=&
\left(
1 - \prod_{ j = 1 }^n S_{a d_j}(\beta-\beta_j)
\right)
F^{\co}_{d_1 \ldots d_n}(\beta_1, \ldots, \beta_n).
\ea

(iv) Bound state residue equation:

For a fusion process $a \times b \rightarrow c$, 
form factors satisfy the bound state residue equation
\be
 -i \lim_{\epsilon \rightarrow 0} \epsilon
F^{\co}_{ab d_1 \ldots a_n}(\beta + i\bar{\theta}_{ac}^b + \epsilon,
\beta - i\bar{\theta}_{bc}^a, \beta_1, \ldots, \beta_n) =
\Gamma_{ab}^c 
F^{\co}_{c d_1 \ldots d_n}(\beta, \beta_1, \ldots, \beta_n),
\ee
where $\bar{\theta}=\pi - \theta$ and $\theta_{ab}^c$ is the fusion
angle. 
Let $n(a, b)=min(a+b, h-a-b)$. 
In the perturbed minimal models, the fusion process occurs for 
$c=n(a, b)$ or $c=|a-b|(\neq 0)$ and the fusion angles are \cite{FKM}
\ba
\theta_{ab}^{|a-b|} &=& ( h - |a-b|) \pi/h, \nn
\theta_{ab}^{n(a,b)} &=& (a+b)\pi/h.
\ea
The on-shell three-point coupling constant $\Gamma_{ab}^c$ is given by
\be
S_{ab}(\beta) \sim 
\frac{ i (\Gamma_{ab}^c)^2 }{\beta - i\theta_{ab}^c }, \qq
{\rm for} \ \beta \sim i\theta_{ab}^c.
\ee
Because the perturbed minimal model is non-unitary, the three point
coupling constant is pure imaginary for the case $c=h-a-b$ ($a+b>N$) 
\cite{FKM}.

The $S$-matrix \Eq{Sab} has a double pole at 
$\beta = (a+b-2c)\pi i/h$ for $c = 1, \ldots, min(a,b)-1$
which corresponds to a weak bound state $ a \times b 
\rightarrow ( (a-c) \times c ) \times ( (b-c) \times c )$
\cite{FKM}. Corresponding to this double pole, 
the form factor has a simple pole at certain rapidity difference. 
We do not give the explicit form of
the (weak) bound state residue equations, which can be found in
\cite{Sm,S}. 

As was shown by Koubek \cite{K}, 
it is sufficient to consider the minimal fusion process 
$a \times b \rightarrow a+b$ ($a+b < N$). 
Informations about the other fusion processes are indirectly contained
in the minimal ones. 

The explicit form of the minimal bound state residue equation is
\bal{bsr}
& & -i \lim_{\epsilon \rightarrow 0} \epsilon
F^{\co}_{ab d_1 \ldots d_n}(\beta + b\pi i/h + \epsilon,
\beta - a\pi i/h, \beta_1, \ldots, \beta_n) \nn
&=&
\Gamma_{ab}^{(a+b)}
F^{\co}_{(a+b) d_1 \ldots d_n}
(\beta, \beta_1, \ldots, \beta_n), \qq a+b \leq N,
\ea
where
\bel
(\Gamma_{ab}^{(a+b)})^2 
= 
2 \tan((a+b)\pi/h)
\frac{ \tan( max(a, b) \pi/h ) }
     { \tan( min(a, b) \pi/h ) }
\prod_{ k = 1 }^{ min(a, b) -1 }
  \left(
    \frac{ \tan( ( max(a, b) + k ) \pi/h ) }
         { \tan( ( min(a, b) - k ) \pi/h ) }
  \right)^2.
\ee
The rest of the bound state residue equations can be derived from
\Eq{bsr}.

(v) Cluster properties \cite{S,KM,MS,DSC}:
\bal{clus}
& & 
\lim_{\Lambda \rightarrow \infty} 
F_{a_1 \ldots a_m a_{m+1} \ldots a_{m+n}}^{\phi_{1, 1+s}}
(\beta_1+\Lambda, \ldots, \beta_m+\Lambda,
\beta_{m+1}, \ldots, \beta_{m+n}) \nn
&=&
\frac{1}{\langle \phi_{1, 1+s} \rangle}
F_{a_1 \ldots a_m}^{\phi_{1, 1+s}}
(\beta_1, \ldots, \beta_m)
F_{a_{m+1} \ldots a_{m+n}}^{\phi_{1, 1+s}}
(\beta_{m+1}, \ldots, \beta_{m+n}).
\ea
Here $\langle \phi_{1, 1+s} \rangle$ is the vacuum expectation value
of the off-critical primary operator $\phi_{1, 1+s}$ \cite{DSC}.
We choose the normalization as follows
\be
\langle \phi_{1, 1+s} \rangle = 1.
\ee

\subsection{Form factors for $\phi_{1, 2}$ and $\phi_{1, 3}$}

As was mentioned in the previous section, the form factors for
$\phi_{1, 2}$ and $\phi_{1, 3}$ are given in the form \Eq{phi12} and
\Eq{phi13} respectively.

The auxiliary objects $f_{\lambda; a_1 \ldots a_n}(\beta_1, \ldots,
\beta_n)$ are defined by
\bal{flambda}
& &
f_{\lambda; a_1\ldots a_n}(\beta_1, \ldots, \beta_n) \nn
&=&
  (-1)^{n-1} 2 \int_{\Gamma_{a_1}(\beta_1)}\frac{d\alpha_1}{2\pi i}\ldots
  \int_{\Gamma_{a_{n-1}}(\beta_{n-1})}\frac{d\alpha_{n-1}}{2\pi i}
  \prod_{i=1}^{n-1}\prod_{j=1}^n
  \varphi_{a_j}(\alpha_i-\beta_j) \nonumber \\
  & &\times\prod_{i<j}^{n - 1}\sinh(\alpha_i-\alpha_j)
  \exp\left(\lambda(\sum_{i=1}^{n-1} \alpha_i
            - \sum_{j=1}^n\beta_j)\right),
  \qq \lambda = 0, \pm 1,
\ea
where $\Gamma_a(\beta)$ is the contour enveloping the points 
$\beta+(a-2l)\pi i/h$, $l=0, 1, \ldots, a$ and
\be
\varphi_a(\beta)=\frac{\displaystyle \prod_{j=1}^{a-1}
 \cosh\frac{1}{2}\left(\beta+(a-2j)\pi i/h\right)}
 {2\displaystyle \prod_{j=0}^a
  \sinh\frac{1}{2}\left(\beta+(a-2j)\pi i/h\right)}.
\ee
Note that our $f_{\lambda; a_1 \ldots a_n}(\beta_1, \ldots, \beta_n)$
corresponds to Smirnov's 
$F_{-\lambda}(\beta_1, \ldots, \beta_n)_{a_1 \ldots a_n}$ \cite{S}.
Although the integration contour $\Gamma_{a_n}(\beta_n)$ is absent in
the expression \Eq{flambda}, all rapidities are on the same footing in
$f_{\lambda; a_1 \ldots a_n}$. See \Eq{flam3}.

The function $\zeta_{ab}(\beta)$ is defined by
\bel{zetaab}
\zeta_{ab}(\beta) = W_{ab}(\beta) F_{ab}^{(min)}(\beta),
\ee
where 
\be
W_{ab}(\beta) = (-1)^{a+\min(a,b)+1}
\frac{\dis 2 \prod_{j=0}^{|a-b|}
\sinh\frac{1}{2}\left(\beta+(|a-b|-2j)\pi i/h\right)}
{\dis \prod_{j=1}^{a+b-1}
\cosh\frac{1}{2}\left(\beta+(a+b-2j)\pi i/h\right)}.
\ee
The phase of $W_{ab}(\beta)$ is chosen such that the cluster equation 
\Eq{clus} holds. 
The minimal two-body form factor $F_{ab}^{(min)}(\beta)$ is given by
\bel{Fabmin}
F_{ab}^{(min)}(\beta) = 
\prod_{ \scriptstyle x = |a-b|+1 
\atop\scriptstyle {\rm step} 2}^{a+b-1}
F_x^{(min)}(\beta).
\ee
Here $F_x^{(min)}(\beta)$ is a building block of the minimal two-body 
form factor:
\bel{Fxmin}
F_x^{(min)}(\beta) = N_x
\exp\left(4\int_0^{\infty}\frac{dk}{k}
\frac{\sin^2(\hat{\beta} k/2\pi)\cosh(1/2-x/h)k
\cosh(k/h)}
{\cosh(k/2)\sinh k}\right),
\ee
where $\hat{\beta} = \pi i -\beta$ and a normalization constant $N_x$
is chosen as 
\bel{Nx}
N_x = \exp
 \left(2\int_0^{\infty}\frac{dk}{k}
  \frac{
   \cosh (k/2) 
   -\cosh \left( 1/2 - x/h\right)k 
    \cosh(k/h)}
    {\cosh(k/2)\sinh k}
 \right).
\ee
$F_x^{(min)}(\beta)$ has no poles or no zeros in the 
strip $0 < Im \beta < 2\pi$. $F_1^{(min)}(\beta)$ has a single
zero at $\beta = 0$.

The constant $\nu_a$ is defined by
\bel{Da}
\nu_a = i^a \left(\frac{2\sin(2a\pi/h)}
{\pi F_{aa}^{(min)}(\pi i)} \right)^{1/2}\ 
\prod_{ l = 1 }^{ a - 1 }\sin(l\pi/h).
\ee
Then the functions \Eq{phi12} and \Eq{phi13} with \Eq{flambda},
\Eq{zetaab} and \Eq{Da} satisfy the form factor bootstrap equations
and are indeed form factors for $\phi_{1,2}$ and $\phi_{1, 3}$
respectively \cite{S}.

In order to transform the form factors into forms suited for summation, 
we rewrite
$f_{\lambda; a_1 \ldots a_n}(\beta_1, \ldots, \beta_n)$ in terms of
$t_i = e^{\alpha_i}$ and $x_j = e^{\beta_j}$. Let 
$\omega = \exp (2\pi i /h) $. 
We have
\bal{flamx}
& & f_{\lambda; a_1\ldots a_n}(\beta_1, \ldots, \beta_n) \nn
&=&
(-1)^{n-1}2^{ n(n+1)/2 } \prod_{j = 1}^n x_j^{ n - \lambda - 1}
\int_{\gamma_{a_1}(x_1)} \frac{dt_1}{2\pi i} \ldots 
\int_{\gamma_{a_{n - 1}}(x_{n-1})} \frac{dt_{n - 1}}{2\pi i} \nn
& & \times
\prod_{i = 1}^{n - 1} \prod_{j = 1}^n \varphi_{a_j}(t_i, x_j)
\prod_{ i < j }^{ n - 1 }
(t_i^2 - t_j^2)
\prod_{ i = 1 }^{ n - 1 } t_i^{ \lambda + 1 },
\ea
where the contour $\gamma_a(x)$ envelops the points 
$x \omega^{ a/2 - l }$ for $l = 0, \ldots, a$.
For $t=e^{\alpha}$ and $x=e^{\beta}$, the function $\varphi_a(t, x)$
is defined by $\varphi_a (\alpha-\beta)=2 t x \varphi_a(t, x)$.
The explicit form of $\varphi_a(t, x)$ is given by
\be
\varphi_a(t, x) = 
\frac{\dis \prod_{j=1}^{a-1} (t + x \omega^{a/2-j})}
{\dis \prod_{j=0}^{a} (t - x \omega^{a/2-j})}.
\ee
With help of a Vandermonde determinant
\be
\prod_{i>j}^{n-1}(t_i^2 - t_j^2)=
{\det}\left( t_i^{2j-2} \right)_{1 \leq i, j \leq n-1},
\ee
we can write \Eq{flamx} in the following form:
\bal{flamx2}
& & f_{\lambda; a_1 \ldots a_n}(\beta_1, \ldots, \beta_n) \nn
&=&
(-1)^{n(n-1)/2} 2^{n(n+1)/2} \prod_{j=1}^n x_j^{n-\lambda-1}
{\det}\left( K^{\lambda}_{a_1 \ldots a_n; ij} 
\right)_{1 \leq i, j \leq n-1},
\ea
where
\bel{Klam}
K^{\lambda}_{a_1 \ldots a_n; ij} = 
\int_{\gamma_{a_i}(x_i)} \frac{ dt}{2\pi i}
t^{2j+\lambda-1} \prod_{k=1}^n \varphi_{a_k}(t, x_k),
\qq
i, j = 1, \ldots, n-1.
\ee
The contour $\gamma_{a_i}(x_i)$ envelops the points 
$t = x_i \omega^{ a_i/2 - l }$ 
for $l = 0, \ldots, a_i$. 

Following the procedure of \cite{S}, we transform the determinant of 
$K^{\lambda}_{a_1 \ldots a_n}$ \Eq{Klam}. 
Let us consider the properties of \Eq{Klam}.
The pole structure of the integrand is determined by 
\be
\prod_{k = 1}^n \varphi_{a_k}(t, x_k).
\ee
The function $\varphi_{a_k}(t, x_k)$ 
($k \neq i$) has no pole in the contour $\gamma_{a_i}(x_i)$.
Thus the value of the integral does not change if 
$\varphi_{a_k}(t, x_k)$ ($k \neq i$) is replaced by
\be
\frac{ t^h - (-1)^{a_k} x_k^h }
{ (-1)^{a_i} x_i^h - (-1)^{a_k} x_k^h }
\varphi_{a_k}(t, x_k).
\ee
Then we have
\ba
& & K^{\lambda}_{a_1 \ldots a_n; ij} \nn
&=&
\int_{\gamma_{a_i}(x_i)} \frac{ dt}{2\pi i}
t^{2j+\lambda-1} \varphi_{a_i}(t, x_i)
\prod_{k\neq i}^n
\frac{\left(t^h - (-1)^{a_k} x_k^h\right)\varphi_{a_k}(t, x_k)}
{(-1)^{a_i} x_i^h - (-1)^{a_k} x_k^h } \nn
&=&
\prod_{k\neq i}^n
\frac{1}{(-1)^{a_i} x_i^h - (-1)^{a_k} x_k^h }
\int_{\gamma_{a_i}(x_i)} \frac{ dt}{2\pi i}
t^{2j+\lambda-1} \prod_{k=1}^n \psi_{a_k}(t, x_k)
\frac{1}{t^h-(-1)^{a_i}x_i^h},
\ea
where
\bal{psifn}
\psi_a(t, x) &=& \left( t^h - (-1)^a x^h \right)
\varphi_a(t, x) \nn
&=&
\prod_{ j = 1 }^{ a - 1 } ( t + x \omega^{a/2 - j})
\prod_{ j = 1 }^{h - a - 1} ( t + x \omega^{(h-a)/2 - j}).
\ea
Now the integrand in the integral over $t$ is regular at the point
$ t = 0 $ and has no singularities except for the points
$x_i \omega^{ a_i/2 - l }$ for $l = 0, \ldots, a_i$.
Thus we can replace the contour $\gamma_{a_i}(x_i)$ by a circle whose
radius is larger than $|x_i|$.
Then we have
\ba
& & {\det}\left( K^{\lambda}_{a_1\ldots a_n; ij} \right)_{1\leq i, j
\leq n-1} \nn
&=&
\prod_{i = 1}^{n - 1} \prod_{j=1 (\neq i)}^n
  \left(
    (-1)^{a_i} x_i^h - (-1)^{a_j} x_j^h
  \right)^{-1} 
{\det}\left( \tilde{K}^{\lambda}_{a_1\ldots a_n; ij} \right)_{1\leq i, j
\leq n-1},
\ea
where
\bel{Klamtil}
\tilde{K}^{\lambda}_{a_1\ldots a_n; ij} \nn
= \oint \frac{dt}{2\pi i}
t^{2j+\lambda-1} \prod_{k=1}^n \psi_{a_k}(t, x_k)
\frac{1}{t^h - (-1)^{a_i} x_i^h}.
\ee
On the contour it holds that $ |t| > | x_j |$. So we can expand 
$( t^h - (-1)^{a_i} x_i^h )^{-1}$ as follows:
\be
\frac{1}{ t^h - (-1)^{a_i} x_i^h }
=
\sum_{q = 1}^{\infty} (-1)^{a_i(q - 1)} x_i^{h(q - 1)}
t^{-h q}.
\ee
Note that after substitution of the above equation into the integral
\Eq{Klamtil}, 
terms with $ q \geq n$ vanish because the highest degree of the
integrand in $t$ is smaller than $-1$. 
The number of non-vanishing terms is at most $n-1$:
\be
\tilde{K}^{\lambda}_{a_1\ldots a_n; ij} \nn
= \sum_{q=1}^{n-1} (-1)^{a_i(q-1)} x_i^{h(q-1)}
\oint \frac{dt}{2\pi i}
t^{2j+\lambda-1-hq} \prod_{k=1}^n \psi_{a_k}(t, x_k).
\ee
The sum over $q$ in the above equation can be interpreted as matrix
product of two matrices of dimension $n-1$. The determinant of 
$\tilde{K}^{\lambda}_{a_1 \ldots a_n}$ becomes product of two
determinants: 
\bal{detK}
& & {\det}\left( \tilde{K}^{\lambda}_{a_1\ldots a_n; ij} 
\right)_{1 \leq i, j \leq n-1} \nn
&=&
{\det}\left( (-1)^{a_i(q-1)} x_i^{h(q-1)} \right)_{1 \leq i, q \leq
n-1}
{\det}\left( 
\oint \frac{dt}{2\pi i}
t^{2j+\lambda-1-hq} \prod_{k=1}^n \psi_{a_k}(t, x_k)
\right)_{1 \leq q, j \leq n-1} \nn
&=&
\prod_{i>j}^{n-1} \left( 
(-1)^{a_i} x_i^h - (-1)^{a_j} x_j^h 
\right)
{\det}\left( 
\oint \frac{dt}{2\pi i}
t^{2j+\lambda-1-hi} \prod_{k=1}^n \psi_{a_k}(t, x_k)
\right)_{1 \leq i, j \leq n-1}.
\ea

To make the meaning of the determinant in \Eq{detK} clear,
it is useful to introduce a notion of ``generalized'' elementary 
symmetric polynomials. 
Recall that the elementary symmetric polynomials with $m$ variables are
defined by 
\be
\prod_{ k = 1 }^m ( t + z_k ) = \sum_{ k \in {\bf Z} } t^{ m - k }
\sigma_k^{(m)}(z_1, \ldots, z_m).
\ee
It holds that $\sigma_k^{(m)} = 0$ if $ k<0, k>m$.
Similarly, let us define generalized elementary symmetric polynomials
by 
\bel{gesp}
\prod_{ k = 1 }^n \psi_{a_k}(t, x_k) =
\sum_{k \in {\bf Z}} t^{ (h - 2) n - k } 
E_{a_1 \ldots a_n; k}(x_1, \ldots, x_n).
\ee
Using the definition of $\psi_a(t, x)$ \Eq{psifn}, we can express
the generalized elementary symmetric polynomial in terms of the
ordinary elementary symmetric polynomials with $(h - 2)n$ variables:
\ba
& & E_{a_1 \ldots a_n; k}(x_1, \ldots, x_n) \nn
&=&
\sigma_k^{((h-2)n)}
\left(\overbrace{
\omega^{a_1/2-1} x_1, \omega^{a_1/2-2} x_1, \ldots, 
\omega^{-a_1/2+1} x_1 }^{a_1 - 1},\right.  \nn
& & \qq \qq \qq
\overbrace{
\omega^{(h-a_1)/2-1} x_1, \omega^{(h-a_1)/2-2} x_1, \ldots, 
\omega^{-(h-a_1)/2+1} x_1}^{h - a_1 - 1 }, \nn
& & \qq \qq \qq \qq
\ldots \ldots \ldots \ldots \ldots \ldots \ldots
\ldots \ldots \ldots \ldots \ldots \ldots \ldots , \nn
& & \qq \qq \qq \qq \qq \qq
\overbrace{
\omega^{a_n/2-1} x_n, \omega^{a_n/2-2} x_n, \ldots, 
\omega^{-a_n/2+1} x_n }^{a_n - 1}, \nn
& & \qq \qq \qq \qq \qq \qq \qq
\left.\overbrace{
\omega^{(h-a_n)/2-1} x_n, \omega^{(h-a_n)/2-2} x_n, \ldots, 
\omega^{-(h-a_n)/2+1} x_n}^{h - a_n - 1}\right). \nonumber
\ea
Note that $E_{a_1 \ldots a_n; k} = 0$ for $ k<0 $ or 
$ k > (h - 2)n $.

For $N=1$, the generalized elementary symmetric polynomials coincide
with the ordinary symmetric polynomials.

The determinant can be written as follows:
\bal{detK2}
& &{\det}\left( 
\oint \frac{dt}{2\pi i}
t^{2j+\lambda-1-hi} \prod_{k=1}^n \psi_{a_k}(t, x_k)
\right)_{1 \leq i, j \leq n-1} \nn
&=&{\det}\left(
E_{a_1 \ldots a_n; h(n - i) - 2(n - j) +\lambda}
\right)_{1 \leq i, j \leq n-1} \nn
&=&
{\det}\left(
E_{a_1 \ldots a_n; hi - 2j +\lambda}
\right)_{1 \leq i, j \leq n-1}.
\ea
Recall that the form factor in the scaling Lee-Yang model ($N=1$, $h=3$)
was proportional to
${\det}( \sigma_{3i-2j+\lambda}^{(n)} )_{1 \leq i, j \leq n-1}$ 
\cite{S,Z}.
Thus, the expression \Eq{detK2} is natural generalization of $N=1$ case.

Then, we have a representation of $f_{\lambda; a_1 \ldots a_n}$:
\bal{flam3}
f_{\lambda; a_1\ldots a_n}(\beta_1, \ldots, \beta_n)
&=&
2^{ n(n + 1)/2}\prod_{j = 1}^n x_j^{ n - \lambda - 1}
\prod_{i > j}^n
  \left(
    (-1)^{a_i} x_i^h - (-1)^{a_j} x_j^h
  \right)^{-1} \nn
& & \qq \times 
{\det}\left( 
E_{a_1 \ldots a_n; hi - 2j +\lambda}
\right)_{1 \leq i, j \leq n-1}, \qq \lambda = 0, \pm 1.
\ea
As was shown in \cite{KS,KO}, in order to represent
two-point correlation function as a Fredholm determinant, 
it is necessary to transform the determinant of the matrix of
dimension $n-1$ into a determinant of a matrix of dimension $n$.

Let us consider the following matrix:
\bel{Mlam}
M^{\lambda}_{a_1 \ldots a_n; ij}(x_1, \ldots, x_n) =
E_{a_1 \ldots a_n; hi-2j+\lambda-h+1}(x_1, \ldots, x_n), \qq 
i, j = 1, \ldots n.
\ee
For $\lambda=0$ or $1$, it holds that
\be
M^{\lambda+1}_{a_1 \ldots a_n; 1j}=
E_{a_1 \ldots a_n; \lambda} \delta_{j, 1}, \qq j=1, \ldots, n,
\ \ \lambda = 0, 1, 
\ee
and $M^{\lambda+1}_{a_1 \ldots a_n; (i+1)(j+1)}=
E_{a_1 \ldots a_n; hi-2j+\lambda}$ for $1 \leq i, j \leq n-1$.
Thus we have
\be
{\det}\left(M^{\lambda+1}_{a_1 \ldots a_n; ij}
\right)_{1 \leq i, j \leq n}
=E_{a_1 \ldots a_n; \lambda}{\det}\left( 
E_{a_1 \ldots a_n; hi - 2j +\lambda}
\right)_{1 \leq i, j \leq n-1}, \qq \lambda= 0, 1.
\ee
For $\lambda = -1$, it holds that
\be
M^{\lambda+h-1}_{a_1 \ldots a_n; nj} = E_{a_1 \ldots a_n ; (h-2)n-1}
\delta_{n, j}, \qq j=1, \ldots, n, \ \ \lambda= -1,
\ee
and $M^{\lambda+h-1}_{a_1 \ldots a_n; ij}=
E_{a_1 \ldots a_n; hi-2j+\lambda}$ for $ 1 \leq i, j \leq n-1$.
Thus we have
\be
{\det}\left(M^{\lambda+h-1}_{a_1 \ldots a_n; ij}
\right)_{1 \leq i, j \leq n}
=E_{a_1 \ldots a_n; (h-2)n-1}{\det}\left( 
E_{a_1 \ldots a_n; hi - 2j +\lambda}
\right)_{1 \leq i, j \leq n-1}, \qq \lambda= -1.
\ee
Note that
\be
E_{a_1 \ldots a_n; 0}=1,
\ee
\be
E_{a_1 \ldots a_n; 1}=
2\cos(\pi/h) \sum_{j=1}^n
\frac{ \sin(a_j\pi/h) }{ \sin(\pi/h) } x_j =
\frac{2\cos(\pi/h)}{m_1} 
\left(\sum_{j=1}^n m_{a_j} e^{\beta_j} \right).
\ee
\be
E_{a_1 \ldots a_n; (h-2)n-1}(x_1, \ldots, x_n)=
\left(\prod_{j=1}^n x_j^{h-2} \right)
E_{a_1 \ldots a_n; 1}(x_1^{-1}, \ldots,
x_n^{-1}).
\ee
Combining the above results with \Eq{phi12} and \Eq{phi13},
the form factors of $\phi_{1, 1+s}$ ($s=1, 2$) can be rewritten as:
\ba
& &F_{a_1 \ldots a_n}^{\phi_{1, 1+s}}(\beta_1, \ldots, \beta_n) \nn
&=&
\tilde{f}_{\lambda; a_1 \ldots a_n}(\beta_1, \ldots, \beta_n)
\left( \prod_{j=1}^n \nu_{a_j} \right) \prod_{i<j}^n
\zeta_{a_i a_j}(\beta_i - \beta_j), \qq s=1, 2,
\ea
where $\lambda = 1$ for $s=1$, and $\lambda = 2$ or $h-2$ for $s=2$ and
\bal{ftillambda}
& & \tilde{f}_{\lambda; a_1 \ldots a_n}(\beta_1, \ldots, \beta_n) \nn
&=& 2^{ n(n + 1)/2}\prod_{j = 1}^n x_j^{ n - \lambda }
\prod_{i > j}^n
  \left(
    (-1)^{a_i} x_i^h - (-1)^{a_j} x_j^h
  \right)^{-1}
{\det}\left( 
M^{\lambda}_{a_1 \ldots a_n; ij}\right)_{1 \leq i, j \leq n}.
\ea
The above auxiliary object has an integral representation similar to 
$f_{\lambda}$ \Eq{flambda}:
\bal{ftillambda2}
\tilde{f}_{\lambda; a_1 \ldots a_n}(\beta_1, \ldots, \beta_n) 
&=&\int_{\Gamma_{a_1}(\beta_1)}\frac{d\alpha_1}{2\pi i}\ldots
  \int_{\Gamma_{a_n}(\beta_n)}\frac{d\alpha_n}{2\pi i}
  \prod_{i=1}^n \prod_{j=1}^n
  \varphi_{a_j}(\alpha_i-\beta_j) \nn
  & &\times\prod_{i<j}^n \sinh(\alpha_i-\alpha_j)
  \exp\left(\lambda \sum_{j=1}^n( \alpha_j - \beta_j)\right).
\ea
In contrast to \Eq{flambda}, 
this expression treats all $\beta_i$ on equal footing.
The equivalence of \Eq{ftillambda2} to \Eq{ftillambda} can be proven
in exactly the same way as for the case of 
$f_{\lambda; a_1 \ldots a_n}$.

\subsection{Form factors for $\phi_{1, 1+s}$ }

Let us analyse properties of \Eq{ftillambda} more closely.

Except for $\lambda = 1, \ldots, 2N$, 
${\det}M^{\lambda}_{a_1 \ldots a_n}$ are trivial:
${\det}M^{\lambda}_{a_1 \ldots a_n}=\delta_{n,0}$.

From the definition of the generalized elementary symmetric
polynomials \Eq{gesp}, we can show that
\be
E_{a_1 \ldots a_n; k}(x_1, \ldots, x_n) = 
\left( \prod_{j=1}^n x_j^{h-2} \right)
E_{a_1 \ldots a_n; (h-2)n-k}(x_1^{-1}, \ldots, x_n^{-1}).
\ee
The matrix $M^{h-\lambda}_{a_1 \ldots a_n}$ is ``isomorphic''
to the matrix $M^{\lambda}_{a_1 \ldots a_n}$ in the sense:
\be
M^{h-\lambda}_{a_1 \ldots a_n; ij}(x_1, \ldots, x_n) =
\left( \prod_{j=1}^n x_j^{h-2} \right)
M^{\lambda}_{a_1 \ldots a_n; (n+1-i)(n+1-j)}
(x_1^{-1}, \ldots, x_n^{-1}).
\ee
Further, it holds that
\be
\left( \prod_{j=1}^n x_j^{-\lambda} \right)
{\det}\left( M^{\lambda}_{a_1 \ldots a_n}(x_1, \ldots, x_n) \right)
=
\left( \prod_{j=1}^n x_j^{\lambda-h} \right)
{\det}\left( M^{h-\lambda}_{a_1 \ldots a_n}(x_1, \ldots, x_n) \right).
\ee
So, we have
\bel{fhll}
\tilde{f}_{h-\lambda; a_1 \ldots a_n}(\beta_1, \ldots, \beta_n) =
\tilde{f}_{\lambda; a_1 \ldots a_n}(\beta_1, \ldots, \beta_n).
\ee
Now, it is easy to guess the form of the form factors for 
the general off-critical primary fields $\phi_{1, 1+s}$
($s=0, \ldots, 2N+1$). 
Suppose that the form factors of $\phi_{1, 1+s}$ is given by 
\bel{FFphis}
F_{a_1 \ldots a_n}^{\phi_{1,1+s}}(\beta_1, \ldots, \beta_n) =
\tilde{f}_{s; a_1 \ldots a_n}(\beta_1, \ldots, \beta_n)
\left( \prod_{j=1}^n \nu_{a_j} \right) \prod_{i<j}^n
\zeta_{a_i a_j}(\beta_i - \beta_j),
\ee
where $\tilde{f}_{s; a_1 \ldots a_n}$ is given by eq.\Eq{ftillambda}
\ba
& & \tilde{f}_{s; a_1 \ldots a_n}(\beta_1, \ldots, \beta_n) \nn
&=& 2^{ n(n + 1)/2}\prod_{j = 1}^n x_j^{ n - s }
\prod_{i > j}^n
  \left(
    (-1)^{a_i} x_i^h - (-1)^{a_j} x_j^h
  \right)^{-1}
{\det}\left( 
M^s_{a_1 \ldots a_n; ij}\right)_{1 \leq i, j \leq n},
\ea
and $M^s_{a_1 \ldots a_n; ij}$ is given by eq.\Eq{Mlam}
\be
M^s_{a_1 \ldots a_n; ij}(x_1, \ldots, x_n) =
E_{a_1 \ldots a_n; hi-2j+s-h+1}(x_1, \ldots, x_n), \qq 
i, j = 1, \ldots n.
\ee
Recall that the definition of the constant $\nu_a$ and the function
$\zeta_{ab}(\beta)$ is given by eq.\Eq{Da} and eq.\Eq{zetaab}
respectively. 

In appendix, we demonstrate that \Eq{FFphis} satisfy the form factor
bootstrap equations. 

The form of the form factor bootstrap equations does not depend on 
the operator. We need to identify the solution with some operator.
The justification of the operator identification in \Eq{FFphis} is the 
following:
From eq.\Eq{fhll}, it holds that $\phi_{1, 1+s}=\phi_{1, 1+h-s}$. 
For $s=0$ or $s=2N+1$ case, the off-critical primary field is the
identity operator and the above form factors give trivial solution.
For $s=1, 2, 2N-1$, eq.\Eq{FFphis} yields the known results \cite{S}.
For general $s$, let us consider the special case of \Eq{FFphis}:
\bel{Fn}
F_n^{\phi_{1, 1+s}}(\beta_1, \ldots, \beta_n):=
F^{\phi_{1, 1+s}}_{1\ldots 1}(\beta_1, \ldots, \beta_n).
\ee
The explicit form of the form
factors of $\phi_{1, 1+2k}$ for $a_1=\ldots=a_n=1$ can be found in
\cite{K}. 
We conjecture that \Eq{Fn} has another equivalent expression:
\bal{Fn2}
& & F_n^{\phi_{1, 1+s}}(\beta_1, \ldots, \beta_n) \nn
&=& (2\nu_1)^n [s]_{\omega^{1/2}}
{\det}\left( [s+2i-2j]_{\omega^{1/2}} \sigma_{2i-j}^{(n)} 
\right)_{1 \leq i, j \leq n-1} \nn
& & \times
\prod_{i<j}^n \frac{ F_{11}^{(min)}(\beta_i-\beta_j)}
{(x_i+x_j) \sinh \frac{1}{2}(\beta_i-\beta_j+2\pi i/h)
\sinh \frac{1}{2}(\beta_i-\beta_j-2\pi i/h)},
\ea
where
\be
[n]_{\omega^{1/2}}=\frac{\omega^{n/2}-\omega^{-n/2}}
{\omega^{1/2}-\omega^{-1/2}} = \frac{\sin(n\pi/h)}{\sin(\pi/h)}.
\ee
We checked that both \Eq{Fn} and \Eq{Fn2} satisfy the same
kinematical residue equations and give the same results for small $n$.
If we set $s=2k$, \Eq{Fn2} agrees with the Koubek's results \cite{K}.
The scaling dimensions of the operators were checked numerically for
small $N$ in \cite{AMV}. These results completely agree with our
operator identification.

Thus the function \Eq{FFphis} gives the form factor for $\phi_{1, 1+s}$.
Eq.\Eq{FFphis} is one of the main results of this paper.

For later convenience, we further rewrite the form factor 
\Eq{FFphis} as follows:
\bel{FFphis2}
F_{a_1 \ldots a_n}^{\phi_{1,1+s}}(\beta_1, \ldots, \beta_n) =
2^n \left( \prod_{j=1}^n \nu_{a_j} x_j^{1-s} \right)
{\det}\left( M^s_{a_1 \ldots a_n} \right)
\prod_{ i<j }^n \frac{ \tilde{\zeta}_{a_i a_j}(\beta_i - \beta_j) }
{ ( x_i x_j )^{(h-2)/2} },
\ee
where
\be
\tilde{\zeta}_{ab}(\beta) = \tilde{W}_{ab}(\beta)
F_{ab}^{(min)}(\beta),
\ee
\be
\frac{2 W_{ab}(\beta - \beta')}{(-1)^b y^h - (-1)^a x^h }
= (x y )^{-h/2} \tilde{W}_{ab}(\beta - \beta'), \qq
x = e^{\beta}, \ y = e^{\beta'}.
\ee
The explicit form of $\tilde{W}_{ab}$ is given by
\be
\tilde{W}_{ab}(\beta) = 
\frac{i^{|a-b|}}{\sinh \frac{1}{2}h(\beta + (a+b)\pi i/h)}
\frac{ \dis 2 \prod_{j=0}^{|a-b|} 
\sinh \frac{1}{2}(\beta + (|a-b|-2j)\pi i/h) }
{\dis \prod_{j=1}^{a+b-1} 
\cosh \frac{1}{2}(\beta +(a+b-2j)\pi i/h)}.
\ee
In the next section, using the expression \Eq{FFphis2}, 
we sum up the two-point correlation 
function \Eq{phiss} into a Fredholm determinant of an integral operator.

\resection{The determinant representation}

A  Fredholm determinant of a linear integral operator $I+V$ can be
decomposed into 
a Taylor series:
\bel{TdefFrdet}
\det(I+V)=\sum_{n=0}^\infty\int\,\frac{dx_1\cdots dx_n}{n!}
{\det}\left( V(x_i, x_j) \right)_{1 \leq i, j \leq n}.
\ee
In order to obtain a determinant representation for the correlation
functions we shall represent the form factor expansion
\Eq{phiss} in the form \Eq{TdefFrdet}. 

The representation of the form factor \Eq{FFphis2} allows us to write
the product of two form factors in \Eq{phiss} as follows:
\ba
& &F_{a_1 \ldots a_n}^{\phi_{1, 1+s}}(\beta_1, \ldots, \beta_n)
F_{a_n \ldots a_1}^{\phi_{1, 1+s'}}(\beta_n, \ldots, \beta_1) \nn
&=&
(-1)^{a_1 + \cdots + a_n} 
\prod_{j=1}^n | 2 \nu_{a_j} |^2 x_j^{2-s-s'}
\prod_{ i<j }^n 
\left|
\frac{ \tilde{\zeta}_{a_i a_j}(\beta_i - \beta_j) }
{ ( x_i x_j )^{(h-2)/2} }
\right|^2 
{\det}\left( M^s_{a_1 \ldots a_n} \right)
{\det}\left( M^{s'}_{a_1 \ldots a_n} \right).
\ea
Note that the phase $(-1)^{a_1 + \cdots + a_n}$ comes from a relation
$\nu_a^2 = (-1)^a | \nu_a |^2$
which is a consequence of non-unitarity.

First, let us consider the product of two determinants.
It is convenient to introduce a matrix:
\bal{Css}
C_{a_1 \ldots a_n; ij}^{(s; s')} &=&
\left( \left(M_{a_1 \ldots a_n}^{s}\right)^T 
M_{a_1 \ldots a_n}^{s'} \right)_{ij} \nn
&=&
\sum_{ k = 1 }^n 
E_{a_1 \ldots a_n; hk-2i + s - h + 1}
E_{a_1 \ldots a_n; hk-2j + s' - h + 1} \nn
&=&
\sum_{ k = -\infty }^n 
E_{a_1 \ldots a_n; hk-2i + s - h + 1}
E_{a_1 \ldots a_n; hk-2j + s' - h + 1}.
\ea
Here we have used the fact that $E_{a_1 \ldots a_n; l}=0$ if $l<0$.

Recall that the generalized symmetric polynomials can be expressed as:
\be
E_{a_1 \ldots a_n; k} = \oint \frac{dt}{2\pi i }
t^{-(h-2)n+k-1} \prod_{l=1}^n \psi_{a_l}(t, x_l),
\ee
where the integration contour is a circle around the origin in
positive direction. 

Substituting the above expression into \Eq{Css} and summing up the
infinite series, we have
\be
C_{a_1 \ldots a_n; jk}^{(s; s')}
= \oint \frac{dt_1}{2\pi i} \oint \frac{dt_2}{2\pi i}
\frac{t_1^{2(n-j)+s}t_2^{2(n-k)+s'}}
{(t_1 t_2)^h -1}
\prod_{l=1}^n \psi_{a_l}(t_1, x_l) \psi_{a_l}(t_2, x_l).
\ee
The radius of the integration contour is chosen to be greater than
one in order for the series to converge.

As in \cite{KS,KO}, let us introduce a matrix 
$D_{a_1 \ldots a_n}^{(s; s')}$ by a linear transformation:
\be
D_{a_1 \ldots a_n}^{(s; s')}= A^T
C_{a_1 \ldots a_n}^{(s; s')} A,
\ee
where
\be
A_{jk} = 
\left. 
   \frac{1}{(n-j)!}\frac{d^{n-j}}{d(x^2)^{n-j}}
   \prod_{l\neq k}^n (x^2 + x_l^2) 
\right|_{x^2=0},
\ee
which has a determinant
\be
{\det} A  = \prod_{ i<j }^n ( x_i^2 -  x_j^2).
\ee
An explicit expression for matrix elements of 
$D_{a_1 \ldots a_n}^{(s; s')}$ is given by
\be
D_{a_1 \ldots a_n;jk}^{(s; s')} = 
\oint \frac{d^2 t}{(2\pi i)^2}
\frac{t_1^s t_2^{s'}}{(t_1 t_2)^h-1}
Y_{a_1 \ldots a_n}(t_1, x_j) Y_{a_1 \ldots a_n}(t_2, x_k),
\ee
where
\be
Y_{a_1 \ldots a_n}(t, x) = 
\frac{J_{a_1 \ldots a_n}(t)}{(t^2 + x^2)},
\ee
and
\be
J_{a_1 \ldots a_n}(t) = 
\prod_{l=1}^n (t^2 + x_l^2) \psi_{a_l}(t, x_l).
\ee
The determinants of matrices
$C_{a_1 \ldots a_n}^{(s; s')}$ and
$D_{a_1 \ldots a_n}^{(s; s')}$ are related by
\be
{\det}\left( C_{a_1 \ldots a_n; jk}^{(s; s')} 
\right)_{1 \leq j, k \leq n} =
{\det}\left( D_{a_1 \ldots a_n; jk}^{(s; s')} 
\right)_{1 \leq j, k \leq n}
\prod_{i<j}^n ( x_i^2 - x_j^2 )^{-2}.
\ee

Now the matrix elements $D_{a_1 \ldots a_n; jk}^{(s; s')}$ 
depend on matrix indices $jk$ through the variables $x_j$ and $x_k$.
However, the element 
$D_{a_1 \ldots a_n;jk}^{(s; s')}$ is not yet a function of only two
arguments, because of $J_{a_1 \ldots a_n}(t)$. This product depends on 
all $x_k$. In order to get rid of this product we introduce an
auxiliary Fock space and auxiliary quantum operators called dual
fields.

Let us define
\bel{Phija}
\Phi_{1a}(x) = q_{1a}(x) + p_{2a}(x), \qq
\Phi_{2a}(x) = q_{2a}(x) + p_{1a}(x), \qq a=1, \ldots, N,
\ee
where the operators $p_{ja}(x)$ and $q_{ja}(x)$ act on the canonical
Fock space in the following way
\be
(0| q_{ja}(x) = 0, \qq p_{ja}(x) |0) = 0, \qq a=1, \ldots, N.
\ee
Non-zero commutators are given by
\be
[p_{1a}(x), q_{1a}(y)] = [ p_{2a}(x), q_{2a}(y) ] = \xi_a(x, y ),
\ee
where
\be
\xi_a(x,y) = \log \left( ( x^2 + y^2 ) 
\psi_a( x, y ) \right).
\ee
Note that in the definitions of $\Phi_{ja}(x)$ \Eq{Phija}, the
coordinate $q_{ja}(x)$ is added to the momentum $p_{(3-j)a}(x)$
conjugate to $q_{(3-j)a}(x)$.

Due to the symmetry of the function 
$\xi_a(x, y) = \xi_a(y, x)$, 
all fields $\Phi_{ja}(x)$ commute with each other
\be
[\Phi_{ja}(x), \Phi_{kb}(y)] = 0, \qq a, b = 1, \ldots, N, \qq
j, k = 1,2.
\ee
Instead of $Y_{a_1 \ldots a_n}(t, x)$ and 
$D_{a_1 \ldots a_n;ij}^{(s; s')}$, 
let us define an operator valued function
\be
\hat{Y}(t, x ) = \frac{ e^{\Phi_1(t)} }{ t^2 + x^2 },
\ee
where
\be
\Phi_1(x) = \sum_{ a = 1 }^n \Phi_{1a}(x),
\ee
and
\bel{hatD}
\hat{D}^{(s; s')}(x, y) = \oint \frac{d^2 t}{(2\pi i)^2}
\frac{t_1^s t_2^{s'}}{(t_1 t_2)^h - 1}
\hat{Y}(t_1, x) \hat{Y}(t_2, y).
\ee
It is easy to show that an exponent of the dual field acts like a
shift operator. Namely, if $g(\Phi_1(y))$ is a function of $\Phi_1(y)$
then 
\be
(0| \left( \prod_{l=1}^n 
e^{\Phi_{2a_l}(x_l)} \right) g\left( \Phi_1(y) \right) |0)
= g \left( \log(J_{a_1 \ldots a_n}(y) )\right).
\ee
Using this property of dual fields one can remove the products
$J_{a_1 \ldots a_n}(t)$ from the matrix
$D_{a_1 \ldots a_n;ij}^{(s; s')}$. For a more detailed derivation one
should look in at formula (3.6) of the paper \cite{KS}. 
Standard arguments of quantum field theory show that
\be
{\det}_n D_{a_1 \ldots a_n}^{(s; s')}
= (0| \left(\prod_{l=1}^n e^{\Phi_{2a_l} (x_l)} \right){\det}
\left( \hat{D}^{(s; s')}
(x_j, x_k) \right)_{1 \leq j,k \leq n} |0).
\ee
Up to now, we have rewritten the polynomial part of the product of
form factors:
\ba
& & F_{a_1 \ldots a_n}^{\phi_{1, 1+s}}(\beta_1, \ldots, \beta_n)
F_{a_n \ldots a_1}^{\phi_{1, 1+s'}}(\beta_n, \ldots, \beta_1) \nn
&=&
(-1)^{a_1 + \cdots + a_n} \prod_{j=1}^n |2 \nu_{a_j}|^2
x_j^{2-s-s'} 
\prod_{i<j}^n
\left| 
\frac{\tilde{\zeta}_{a_i a_j}(\beta_i-\beta_j)}
{(x_i x_j)^{(h-2)/2}(x_i^2-x_j^2)}\right|^2 \nn
& & \qq \times
{\det}\left( D_{a_1 \ldots a_n; jk}^{(s; s')} 
\right)_{1 \leq j, k \leq n}.
\ea
In order to factorize the double product part, we introduce 
another set of dual fields
\be
\tilde{\Phi}_{0a}(x) = \tilde{q}_{0a}(x) + \tilde{p}_{0a}(x),
\qq a=1, \ldots, N.
\ee
As usual
\be
(0|\tilde{q}_{0a}(x)=0, \qq 
\tilde{p}_{0a}(x)|0)=0, \qq a=1, \ldots, N.
\ee
The operators $\tilde{q}_{0a}(x)$ and $\tilde{p}_{0a}(x)$ commute with 
all $q_{ja}(x)$ and $p_{ja}(x)$ ($j=1, 2$ and $a=1, \ldots N$). 
The non-zero commutators are given by
\be
[\tilde{p}_{0a}(x), \tilde{q}_{0b}(y)] = 
\eta_{ab}(x,y) = \eta_{ba}(x,y),
\ee
where
\be
\eta_{ab}(x,y) = \eta_{ab}(y, x) = 2 \log \left|
\frac{\tilde{\zeta}_{ab}(\log(x/y))}{(x y)^{(h-2)/2}(x^2 - y^2)}
\right|.
\ee
Because $|\zeta_{ab}(\beta)|$ is a symmetric
function $|\zeta_{ab}(-\beta)|=|\zeta_{ba}(\beta)|$,
the newly introduced dual fields also mutually commute
\bel{DAbel0}
[\tilde\Phi_{0a}(x),\tilde\Phi_{0b}(y)] = 0 =
[\tilde\Phi_{0a}(x),\Phi_{jb}(y)].
\ee

Due to the fact that
$\zeta_{aa}(\beta)$ has a zero of first order at $\beta=0$, 
$\eta_{aa}(x, y)$ has no singularity at $x=y$.
\be
\eta_{aa}(x, x) = -2 \log | \lambda_a x^h |,
\ee
where
\be
\lambda_a = \frac{h}{|\nu_a|^2}.
\ee
Here we used the value of the derivative of $\zeta_{aa}(\beta)$ at
$\beta=0$:
\be
|\tilde{\zeta}_{aa}'(0)| = \frac{2}{h} | \nu_a |^2.
\ee
The Campbell-Hausdorff formula yields
\be
(0| \prod_{l=1}^n e^{\tilde{\Phi}_{0a_l}(x_l)} |0) = 
\prod_{i=1}^n \prod_{j=1}^n e^{\frac{1}{2}\eta_{a_i a_j}(x_i, x_j)}
= \prod_{j=1}^n \lambda_{a_j}^{-1} x_j^{-h} \prod_{i<j}
\left|
\frac{\tilde{\zeta}_{a_i a_j}(\log(x_i/x_j))}
{(x_i x_j)^{(h-2)/2} (x_i^2 - x_j^2)} 
\right|^2.
\ee
Combining these results, we can represent the product of two form
factors in \Eq{phiss} as follows:
\ba
& & F_{a_1 \ldots a_n}^{\phi_{1, 1+s}}(\beta_1, \ldots, \beta_n)
F_{a_n \ldots a_1}^{\phi_{1, 1+s'}}(\beta_n, \ldots, \beta_1) \nn
&=&
(-1)^{a_1 + \cdots + a_n} (4h)^n \nn
& & \qq \times
(0| \prod_{l=1}^n e^{\Phi_{0a_l}(x_l)}
{\det}\left( (x_j x_k)^{(h-s-s' + 2)/2}
\hat{D}^{(s; s')}(x_j, x_k) \right)_{1 \leq j, k \leq n}
|0),
\ea
where
\be
\Phi_{0a}(x) = \tilde{\Phi}_{0a}(x) + \Phi_{2a}(x).
\ee
Then two point function \Eq{phiss} can be written as
\ba
& & \langle \phi_{1, 1+s}(x) \phi_{1, 1+s'}(0) \rangle \nn
&=& \sum_{n=0}^{\infty} \frac{1}{n!} \left( \frac{2h}{\pi} \right)^n
\int_0^{\infty} d^n x \nn
& & \qq \times
(0| \left(\prod_{l=1}^n V(x_l)\right)
 {\det}\left( (x_j x_k)^{(h-s-s' + 1)/2}
\hat{D}^{(s; s')}(x_j, x_k)
\right)_{1 \leq j, k \leq n} |0),
\ea
where
\be
V(x)=  \sum_{a=1}^N (-1)^a e^{\Phi_{0a}(x) - \theta_a(x)},
\ee
with $\theta_a(x) = r m_a (x + x^{-1})/2$.
Because all operators $\Phi_{0a}(x)$ commute 
with each other, we can formally
define the logarithm of the operator $V(x)$ by
\be
e^{\Phi_0(x)} = V(x) = \sum_{a=1}^N (-1)^a e^{\Phi_{0a}(x) - \theta_a(x)}.
\ee
Finally we obtain a determinant representation in terms of an
integral operator
\bal{detss}
& & \langle \phi_{1, 1+s}(x) \phi_{1, 1+s'}(0) \rangle \nn
&=& \sum_{n=0}^{\infty} \frac{1}{n!} \left( \frac{2h}{\pi} \right)^n
\int_0^{\infty} d^n x \nn
& & \qq \times
(0| {\det}\left( (x_j x_k)^{(h-s-s' + 1)/2}
\hat{D}^{(s; s')}(x_j, x_k)
e^{\frac{1}{2} \Phi_0(x_j) + \frac{1}{2} \Phi_0(x_k) }
\right)_{1 \leq j, k \leq n} |0) \nn
&=&(0| {\det}\left( I + \hat{U}^{(s; s')} \right) |0),
\ea
where an integral operator is defined by
\be
U^{(s; s')}(x, y)
= 2h \pi^{-1}
(x y )^{(h-s-s' + 1)/2}
\hat{D}^{(s; s')}(x, y) 
e^{\frac{1}{2} \Phi_0(x) + \frac{1}{2} \Phi_0(y) }.
\ee
Here the integral operator $\hat{D}^{(s;s')}$ is given by \Eq{hatD}.

\resection{Discussion}

In this paper, we have considered the two-point correlation functions
in the perturbed minimal models $M_{2, 2N+3}+\phi_{1, 3}$. 

It is known that the operator content of
the perturbed model is same as the unperturbed models \cite{K}. 
The model contains $N+1$ off-critical primary fields $\phi_{1, 1+s}$.

We have determined the explicit form of the form factors for the
off-critical primary fields $\phi_{1, 1+s}$ \Eq{FFphis}.
The information about the operator $\phi_{1, 1+s}$ is carried by the
function $\tilde{f}_{s; a_1 \ldots a_n}$ \Eq{ftillambda2}:
\ba
\tilde{f}_{s; a_1 \ldots a_n}(\beta_1, \ldots, \beta_n) 
&=&\int_{\Gamma_{a_1}(\beta_1)}\frac{d\alpha_1}{2\pi i}\ldots
  \int_{\Gamma_{a_n}(\beta_n)}\frac{d\alpha_n}{2\pi i}
  \prod_{i=1}^n \prod_{j=1}^n
  \varphi_{a_j}(\alpha_i-\beta_j) \nn
  & &\times\prod_{i<j}^n \sinh(\alpha_i-\alpha_j)
  \exp\left(s \sum_{j=1}^n( \alpha_j - \beta_j)\right).
\ea
This representation reveals remarkably simple structure of the
operator content of the perturbed minimal model.

Recall that 
the perturbed minimal model can be described as the restriction of the 
sine-Gordon model at the coupling constant $g^2/8\pi = 2/(2N+3)$ \cite{S}.
In the restricted sine-Gordon model, the off-critical primary field
$\phi_{1, 1+s}$ corresponds to the following exponential operator:
\be
{\cal P} e^{i s g \phi /2 } {\cal P},
\ee
where $\phi$ is the sine-Gordon field and 
${\cal P}$ is the projection operator into the soliton-free
sector \cite{S,K}. 
If we use the representation 
\Eq{ftillambda2} (not \Eq{ftillambda}) and replace $2\pi/h$ by 
$\xi=\pi g^2/(8\pi - g^2)$, the form factor \Eq{FFphis} becomes the
breather form factor for the exponential operator $e^{i s g \phi /2 }$ 
in the unrestricted sine-Gordon model.
The expression \Eq{Fn2} remains valid by this replacement and it gives 
the form factors for the lightest
breathers. It can be obtained from the form factors of the exponential 
operator in the sinh-Gordon model by analytic continuation in the
coupling constant \cite{Luk}.

Using a representation of the form factor \Eq{FFphis2}, we have
obtained determinant representation for two-point correlation function
of off-critical primary fields \Eq{detss}, which is natural
generalization of that of the scaling Lee-Yang model \cite{KO}.

It would be very interesting if one could extract some non-perturbative 
features from the determinant representations \Eq{detss}.

%%%%%%%%%%%%%%%%%%%%%%%%%%%%%%%%%%%%%%%%%%%%%%
\section*{Acknowledgments}

I would like to thank Prof. R. Sasaki for careful reading of the
manuscript. 
%%%%%%%%%%%%%%%%%%%%%%%%%%%%%%%%%%%%%%%%%%%%%%%%%%%%%%%%%%%%%%

\appendix

\resection{Appendix}

In this appendix, we collect some relations which are helpful to show
that the function \Eq{FFphis} satisfies the
form factor bootstrap equations (i)-(v).

There is no difficulty in proving (i) Watson's equation and 
(ii) Lorentz covariance.

Note that the minimal building block of the two-body form factor 
$F_x^{(min)}(\beta)$ \Eq{Fxmin} has a property
\be
F_x^{(min)}(\beta) = \{ x \}_{(\beta)} F_x^{(min)}(-\beta),
\ee
\be
F_x^{(min)}(\beta + 2\pi i) = F_x^{(min)}(-\beta).
\ee
Then the minimal two-body form factor $F_{ab}^{(min)}(\beta)$ 
\Eq{Fabmin} satisfies Watson's equation for $n=2$:
\be
F_{ab}^{(min)}(\beta) = S_{ab}(\beta) F_{ab}^{(min)}(-\beta),
\ee
\be
F_{ab}^{(min)}(\beta+2\pi i) = F_{ab}^{(min)}(-\beta).
\ee
Using these relations, one can easily check that \Eq{FFphis} obeys 
Watson's equations for general $n$. 

(iii) Kinematical residue equation:

To show that \Eq{FFphis} satisfies the kinematical residue equation, 
we need the following relations.

The residue of $\zeta_{aa}(\beta)$ at $\beta = \pi i$ is given by
\be
-i \lim_{\epsilon \rightarrow 0} \epsilon \zeta_{aa}(\pi i + \epsilon)=
(-1)^{a-1} 4i F_{aa}^{(min)}(\pi i) \prod_{j=1}^{a-1}\sin^{-2}(j\pi/h).
\ee

Using a representation of $\tilde{f}_{\lambda}$ \Eq{ftillambda2}, one
can show that
\ba
& & \tilde{f}_{ \lambda; a a d_1 \ldots d_n }
      ( \beta + \pi i, \beta, \beta_1, \ldots, \beta_n ) \nn
&=&
\frac{i}{\sin(2\pi a/h)}
\left[
\prod_{ j = 1 }^n
\varphi_{d_j}(\beta - \beta_j + \pi i - a\pi i/h)
\varphi_{d_j}(\beta - \beta_j + a\pi i/h) \right.\nn
& & \qq -
\left.
\prod_{ j = 1 }^n
\varphi_{d_j}(\beta - \beta_j + \pi i + a\pi i/h)
\varphi_{d_j}(\beta - \beta_j - a\pi i/h) \right] \nn
& & \qq \qq \times
\tilde{f}_{ \lambda; d_1 \ldots d_n } ( \beta_1, \ldots, \beta_n ).
\ea

With the choice of the normalization \Eq{Nx}, the building block
\Eq{Fxmin}
satisfies 
\ba
& & F_x^{(min)}(\beta + \pi i ) F_x^{(min)}(\beta) \nn
&=& - \cosh \frac{1}{2}(\beta - (x-1)\pi i/h )
\cosh \frac{1}{2}(\beta - (x+1)\pi i/h ) \nn
& & \qq \times \sinh \frac{1}{2}(\beta + (x-1)\pi i/h )
\sinh \frac{1}{2}(\beta + (x+1)\pi i/h).
\ea
Using this relation, we can show that
\ba
& &\zeta_{ad}(\beta+\pi i) \zeta_{ad}(\beta) \nn 
&=& 
\varphi_{a}^{-1}(\beta+ d\pi i/h)
\varphi_{a}^{-1}(\beta+\pi i - d\pi i/h) \nn
&=& 
\varphi_d^{-1}(\beta+ a\pi i/h)
\varphi_d^{-1}(\beta+\pi i - a \pi i /h).
\ea

It holds that
\be
\frac{
\varphi_d(\beta + \pi i + a\pi i/h)
\varphi_d(\beta - a\pi i/h) 
}{
\varphi_d(\beta + \pi i - a\pi i/h)
\varphi_d(\beta + a\pi i/h)
} = S_{ad}(\beta).
\ee

Making use of these relations, we can show 
the function \Eq{FFphis} satisfies
the kinematical residue equations.

(iv) Bound state residue equation:

In order to verify that \Eq{FFphis} satisfies the bound state residue
equation for the minimal fusion process 
$a \times b \rightarrow (a+b)$ ($a+b\leq N$), 
we need the following relations.

\ba
& & -i \lim_{\epsilon \rightarrow 0} \epsilon
\tilde{f}_{\lambda; ab d_1 \ldots d_n}(\beta + b\pi i/h + \epsilon,
\beta - a\pi i/h, \beta_1, \ldots, \beta_n) \nn
&=&
(-1)^n i^{a-b+1} \mu_a \mu_b 
\tilde{f}_{\lambda; (a+b) d_1 \ldots d_n}
(\beta, \beta_1, \ldots, \beta_n)
\prod_{j = 1}^n \varphi_{d_j}(\beta - \beta_j + (b-a)\pi i/h ),
\ea
where
\be
\mu_a =  i^{-a} \oint \frac{d\alpha}{2\pi i} 
\varphi_a(\alpha - a\pi i/h)
= \frac{\dis \prod_{j=1}^{a-1} \cos(j\pi/h)}
{\dis \prod_{j=1}^a \sin(j\pi/h)}.
\ee
It holds that
\be
\oint \frac{d\alpha}{2\pi i} \varphi_{(a+b)}(\alpha) =
i^{a-b} \mu_a \mu_b =
\oint \frac{d\alpha}{2\pi i} \varphi_a(\alpha-a\pi i/h)
\oint \frac{d\alpha'}{2\pi i} \varphi_b(\alpha' +b\pi i/h).
\ee
The function $\zeta_{ab}(\beta)$ satisfies a bootstrap equation:
\be
\zeta_{ad}(\beta + b\pi i/h) \zeta_{bd}(\beta - a\pi i/h) =
- \varphi_{d}^{-1}(\beta + (b-a)\pi i/h)
\zeta_{(a+b)d}(\beta).
\ee
There is a relation among constants:
\be
\frac{F_{(a+b)(a+b)}^{(min)}(\pi i)}
{F_{aa}^{(min)}(\pi i)F_{bb}^{(min)}(\pi i)}
\left(
F_{ab}^{(min)}((a+b)\pi i/h)
\right)^2 = 
\prod_{\scriptstyle x = 2max(a,b)+1 
\atop\scriptstyle {\rm step} \ 2}^{2a+2b-1}
\sin^2\frac{(x-1)\pi}{2h} \sin^2\frac{(x+1)\pi}{2h},
\ee
\be
\frac{2 \dis \prod_{j=min(a,b)}^{max(a,b)} \sin(j\pi/h) }
{\dis \prod_{j=1}^{a+b-1} \cos(j\pi/h) }
\frac{\nu_a \nu_b }
{\nu_{(a+b)}} \mu_a \mu_b F^{(min)}_{ab}((a+b)\pi i/h)
=\Gamma_{ab}^{(a+b)}.
\ee

With the aid of these relations, one can prove that the function
\Eq{FFphis} satisfy bound state residue equations.

(v) Cluster properties:

Finally we analyse cluster properties \Eq{clus}.

For $\beta \rightarrow \pm \infty$, the building block of the minimal
two-body form factor behaves as
\be
F_x^{(min)}(\beta) = - \frac{1}{4} e^{|\beta|} + \ldots.
\ee
Therefore, for $\beta \rightarrow \infty$,
\be
\zeta_{ab}(\beta) = (-1)^{a+1} \frac{1}{2} e^{\beta} + \ldots,
\ee
and for $\beta \rightarrow - \infty$,
\be
\zeta_{ab}(\beta) = (-1)^b \frac{1}{2} e^{-\beta} + \ldots.
\ee
Let us consider the large $\Lambda$ limit of 
\be
M^{\lambda}_{a_1 \ldots a_m a_{m+1} \ldots a_{m+n}}
(e^{\Lambda} x_1, \ldots, e^{\Lambda} x_m, x_{m+1}, \ldots, x_{m+n}).
\ee
Similar to the case of the ordinary elementary symmetric polynomials 
\cite{KM,MS}, the leading behavior of the generalized symmetric
polynomials is determined by the highest degree term.

If $k \leq (h-2)m$,
\ba
& & E_{a_1 \ldots a_m a_{m+1} \ldots a_{m+n}; k}
(e^{\Lambda} x_1, \ldots, e^{\Lambda} x_m, x_{m+1}, \ldots, x_{m+n})
\nn
&\sim&
e^{k\Lambda}  E_{a_1 \ldots a_m; k}(x_1, \ldots, x_m).
\ea
If $k> (h-2)m$,
\ba
& & E_{a_1 \ldots a_m a_{m+1} \ldots a_{m+n}; k}
(e^{\Lambda} x_1, \ldots, e^{\Lambda} x_m, x_{m+1}, \ldots, x_{m+n})
\nn
&\sim&
e^{(h-2)m\Lambda}  E_{a_1 \ldots a_m; (h-2)m}(x_1, \ldots, x_m)
E_{a_{m+1} \ldots a_{m+n}; k-(h-2)m }(x_{m+1}, \ldots, x_{m+n}) \nn
&=&
e^{(h-2)m\Lambda} \left(\prod_{j=1}^m x_j^{h-2}\right)
E_{a_{m+1} \ldots a_{m+n}; k-(h-2)m }(x_{m+1}, \ldots, x_{m+n}).
\ea

Then
\ba
& & {\det} \left(
M^{\lambda}_{a_1 \ldots a_m a_{m+1} \ldots a_{m+n}}
(e^{\Lambda} x_1, \ldots, e^{\Lambda} x_m, x_{m+1}, \ldots, x_{m+n})
\right) \nn
&=&
{\det} \left(
E_{a_1 \ldots a_m a_{m+1} \ldots a_{m+n}; hi-2j+\lambda-h+1}
(e^{\Lambda} x_1, \ldots, e^{\Lambda} x_m, x_{m+1}, \ldots, x_{m+n})
\right)_{1 \leq i, j \leq m+n} \nn
&\sim&
{\det} \left(
E_{a_1 \ldots a_m a_{m+1} \ldots a_{m+n}; hi-2j+\lambda-h+1}
(e^{\Lambda} x_1, \ldots, e^{\Lambda} x_m)
\right)_{1 \leq i, j \leq m} \times
e^{(h-2)mn\Lambda} \left(\prod_{j=1}^m x_j^{(h-2)n}\right) \nn
& & \qq \times
{\det} \left(
E_{a_{m+1} \ldots a_{m+n}; h(i-m)-2(j-m)+\lambda-h+1}
(x_{m+1}, \ldots, x_{m+n})
\right)_{m+1 \leq i, j \leq m+n} \nn
&=&
\exp\left[\left((h-2)(n+\frac{m-1}{2})+\lambda \right)m \Lambda
\right] \left(\prod_{j=1}^m x_j^{(h-2)n}\right) \nn
& & \qq \times {\det} \left(
M^{\lambda}_{a_1 \ldots a_m}
(x_1, \ldots, x_m )\right) 
{\det} \left(
M^{\lambda}_{a_{m+1} \ldots a_{m+n}}
(x_{m+1}, \ldots, x_{m+n} )\right).
\ea
These results allow us to verify that the functions \Eq{FFphis}
satisfy cluster equation with normalization 
$\langle \phi_{1, 1+s} \rangle =1$.

\end{document}